\documentclass[fleqn,10pt]{wlscirep}
\usepackage[utf8]{inputenc}
\usepackage[T1]{fontenc}
\usepackage{bm}
\usepackage{graphicx}
\usepackage{braket}
\usepackage{amssymb}
\usepackage{textcomp}
\usepackage{lineno}
\usepackage{xcolor}
\usepackage{soul}
\usepackage{xspace}
\usepackage{circledsteps}

\title{Quantum evolution: terrestrial fine-tuning of magnetic parameters} 
\author[*1,2,3,5]{Betony Adams}
\author[*1,2,3]{Abbas Hassasfar}
\author[3,4]{Ilya Sinayskiy}
\author[5,6]{Alistair Nunn}
\author[5]{Geoffrey Guy}
\author[1,2,3]{Francesco Petruccione}
\affil[1]{School of Data Science and Computational Thinking, Stellenbosch University, Stellenbosch 7602, South Africa}
\affil[2]{School of Physics, Stellenbosch University, Stellenbosch 7602, South Africa}
\affil[3]{National Institute for Theoretical and Computational Sciences (NITheCS), South Africa}
\affil[4]{School of Chemistry and Physics, University of KwaZulu-Natal, Durban, South Africa}
\affil[5]{The Guy Foundation, Dorset, UK}
\affil[6]{Research Centre for Optimal Health, University of Westminster, London, UK}
\affil[*] {corresponding authors: betony@gmail.com and 27522644@sun.ac.za}

\date{\today}

\begin{abstract}
For the first time in history, humankind might conceivably begin to imagine itself as a multi-planetary species. This goal will entail technical innovation in a number of contexts, including that of healthcare. All life on Earth shares an evolution that is coupled to specific environmental conditions, including gravitational and magnetic fields. While the human body may be able to adjust to short term disruption of these fields during space flights, any long term settlement would have to take into consideration the effects that different fields will have on biological systems, within the space of one lifetime, but also across generations. Magnetic fields, for example, influence the growth of stem cells in regenerative processes. Circadian rhythms are profoundly influenced by magnetic fields, a fact that will likely have an effect on mental as well as physical health. Even the brain responds to small perturbations of this field. One possible mechanism for the effects of weak magnetic fields on biological systems has been suggested to be the radical pair mechanism. The radical pair mechanism originated in the context of spin chemistry to describe how magnetic fields influence the yields of chemical reactions. This mechanism was subsequently incorporated into the field of quantum biology. Quantum biology, most generally, is the study of whether non-trivial quantum effects play any meaningful role in biological systems. The radical pair mechanism has been used most consistently in this context to describe the avian compass. Recently, however, a number of studies have investigated other biological contexts in which the radical pair might play a role, from the action of anaesthetics and antidepressants, to microtubule development and the proper function of the circadian clock. With this in mind, we investigated how magnetic fields ranging in strengths comparable to those of the planets in our own solar system would influence the dynamics of the radical pair, with possible knock on health effects. We also investigate what light this may shed on the development and evolution of biological materials in Earth's magnetic field. Our results demonstrate that radical pair yields can be tuned to the Earth's magnetic field through the intrinsic magnetic properties of biological materials. We conclude that further attention should be paid to biological responses in magnetic fields and particularly to the development of technologies and therapeutics which might properly investigate and exploit these effects.
\end{abstract}

\begin{document}
\flushbottom
\maketitle

\thispagestyle{empty}

\section{Introduction}
The history of human settlement is intimately linked to the history of medicine. New worlds mean new diseases and expansion beyond Earth poses an entirely new set of health problems. The human body has evolved over millions of years to a very specific electromagnetic and gravitational environment, where our physiology is coupled to the Earth’s rhythms and characteristics, one of which is the geomagnetic field. If we are to thrive under conditions of different magnetic fields, it seems important we understand exactly the effect our physical environment -- and disruption to this environment -- may have on biological systems. It is already acknowledged that the geomagnetic field plays a role in protecting and facilitating life on Earth by deflecting harmful cosmic rays, among other things \cite{erdmann}. What is less well known, however, is that weak, Earth-strength magnetic fields can have a more direct effect on biological systems, potentially through the modulation of quantum spin in the molecules that constitute these systems \cite{simon}. Spin is the property of quantum particles such as electrons and protons that describes their response to a magnetic field. Spin has already played a role in revolutionising medical techniques through the invention of magnetic resonance imaging (MRI). The difference is that MRI machines exploit the spin-polarising effects of strong magnetic fields \cite{ai}. What is still not yet properly understood is exactly how weak magnetic fields, through mechanisms such as radical pairs, are implicated in biological and physiological – and, indeed, even psychological – outcomes \cite{simon}.\\
\\
The experimental evidence is suggestive. A number of fundamental biological processes are influenced by weak magnetic fields, where by `weak' we mean Earth strength and smaller (in the range micro- to nanotesla). Circadian rhythms, for instance, are sensitive to magnetic fields \cite{yoshi,xue}. Stem cell development, both neurogenesis and other, is modulated by weak magnetic fields; a phenomenon that is thought to be mediated by reactive oxygen species (ROS) \cite{zhang1,zhang2,beane,vanhuizen}. Magnetic fields have also been shown to be important to a number of different biological functions with further implications. Birds can sense the geomagnetic field and potentially use it as a compass in their migrations \cite{hore2016,rodgers,cai,pauls}. But experiments show that the human brain also responds to perturbations of the geomagnetic field. Magnetic field changes result in a decrease in amplitude of alpha frequency (8-13 Hz) brain waves, which are related to the brain's processing of external stimuli \cite{wang}. Research also suggests that disruptions to the Earth’s field caused by geomagnetic storms correlate with physiological and psychological changes, including increased instances of suicide (though it is unclear whether the effect is due to direct magnetic effects or increased solar radiation) \cite{close,kay,ora}. Magnetic fields, it would seem, are integral to both physical and mental health. However, our understanding of the mechanism – our ability to predict, exploit and drive these effects – is not well developed. The field is expanding theoretically, with the radical pair model being applied to a number of diverse contexts \cite{simon,simon1,simon2,simon3,simon4}. But what is still needed is a better understanding of the functional relevance of quantum spin in vivo and the exact parameters of this magnetic interaction.\\
\\
In this paper we first introduce to what extent magnetic fields are implicated in the functioning of biological organisms, with a short overview of quantum biology. We then present a demonstration of the effects of the different magnetic fields on a physiologically relevant outcome such as ROS. In the first part of the paper we illustrate, with a simple radical pair model, the differing effects of magnetic field strengths corresponding to the planets in our solar system. This serves as a proof of concept: that settlement on planets with different magnetic fields requires further investigation of the ways in which magnetic fields interact with biological organisms. In the second part of the paper, we explore in more detail the relationship between spin state probability and environmental parameters dictated, for example, by the hyperfine or Zeeman effect. We do this with the aim of illustrating how the magnetic properties of living materials may have evolved to be fine-tuned to their terrestrial environment.\\
\\
\subsection{Quantum biology in brief}
Quantum biology is commonly referred to as an emerging field of research. In reality, however, the development of quantum physics often spurred questions about how the theory might illuminate those mysteries at the heart of living systems \cite{mcfadden}. Many of the founding figures of quantum physics elaborated on these questions \cite{bohr,schrodinger}. Definitions of quantum biology differ, but it might be said that it encompasses the study of quantum effects in the remarkable materials that constitute biological systems \cite{adams2023}. Indeed, where quantum physics emerged out of the necessity for a new theory to describe new observations with regards to the interactions of light and matter, quantum biology is equally interested in the investigation of how light interacts with living matter. Light-matter interactions in biological systems are a growing field of interest in a number of contexts. Photo-activation drives energy and charge transfer events in processes such as photosynthesis. Photosynthesis involves the movement of energy along electron transport chains, for the purpose of adenosine triphosphate (ATP) synthesis. Energy is transported between light sensitive molecules known as chromophores, which are embedded in proteins. There has been a great deal of investigation into the question of whether this energy transport utilises quantum coherence \cite{engel,brixner,vangrondelle,schlau,panit,collini}. While there is yet to be a definitive answer to this question, chromophores are ubiquitous in biological materials; and energy and charge transfer processes in photosynthesis are equally relevant to a number of other important biological contexts, including the mitochondrial electron transport chain central to metabolic processes and respiration. Quantum biology encompasses a wide range of topics in addition to photosynthesis: enzyme catalysis \cite{devault,klinman}, DNA mutation \cite{lowdin,slocombe}, receptor binding \cite{turin96,turinVTO,hoehn}, microtubule and mitochondrial function \cite{craddock14,craddockbeats,kurian,celardo}, magnetoreception \cite{schulten78,mouritsen2005,ritz2000,horerodgers}, regulation of the production of ROS \cite{usselman,usselman2}, calcium ion storage and release \cite{fisher,fisher2,fisher3} and, potentially, consciousness \cite{hameroff2014}. For the purposes of this paper we will focus only on one of the many topics of quantum biology: the spin dynamics of radical pairs, most specifically in the context of ROS. However, while the functional biological context may be specific, the fundamental quantum effects more generally involve the behaviour -- whether we are looking at charge or spin -- of particles such as electrons, protons, neutrons and phonons.\\ 
\\
\subsection{Spin chemistry and the radical pair mechanism}
Whereas transfer processes in photosynthesis and other contexts are largely preoccupied with the property of charge, the radical pair mechanism focuses on spin. Spin is an intrinsic property of matter in the same way that mass and charge are intrinsic properties of matter. Spin describes how matter behaves in a magnetic field. The radical pair mechanism describes how the spins of paired electrons behave in a magnetic field, where a radical is a molecule that has an unpaired electron. The radical pair mechanism was first developed as a model in the context of spin chemistry, as a way to describe how magnetic fields change the yields of chemical reactions \cite{steiner,horespin}. A radical pair consists of two unpaired electrons that are spatially separated but spin correlated. This spin correlation can take the form of a singlet state or one of three triplet states. These states describe different alignments of the spins of each electron with respect to the other. Once the radical pair has been formed it undergoes singlet-triplet oscillation due to the interaction of the electron spins with environmental factors such as the external magnetic field (Zeeman effect) and the magnetic moments of the surrounding nuclei in the molecule (hyperfine effect). Singlet and triplet states result in different chemical outcomes. The ratio of these spin states can be altered by the application of magnetic fields. In this way magnetic fields can alter the outcomes of chemical reactions \cite{steiner,horerodgers,horespin}. The mechanism is described in some detail in Figure 1. At its simplest, however, the radical pair mechanism depends on two main principles: that magnetic fields can modulate spin states and that spin states can modulate the outcome of chemical reactions. \\
\begin{figure}
	\includegraphics[width=\textwidth]{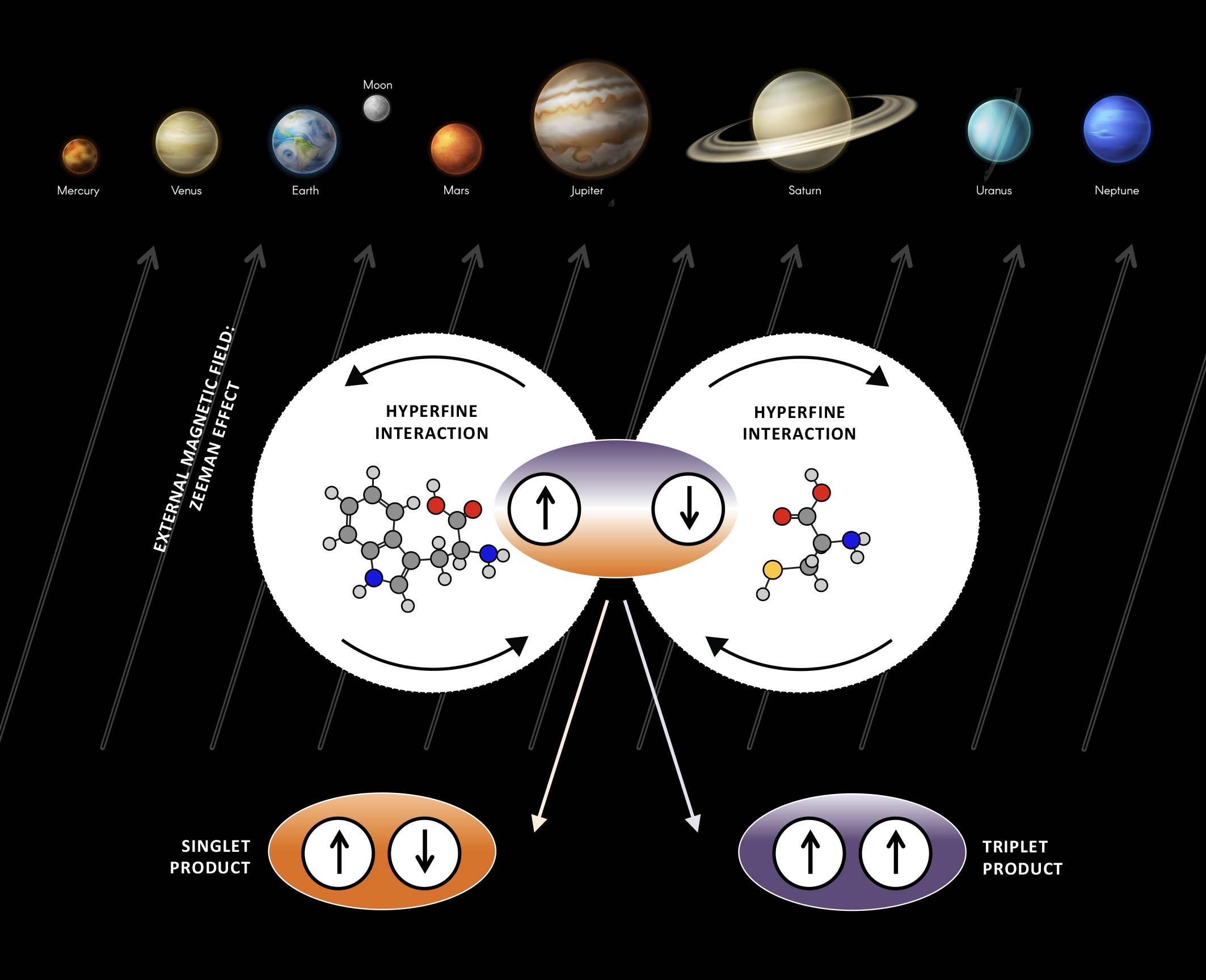}
	\caption{An illustration of the radical pair mechanism, with electrons of the pair given by \textcircled{$\uparrow$} and \textcircled{$\downarrow$} to represent their spin. The creation of the radical pair involves an electron transfer facilitated either by light or chemical reaction, resulting in a pair of electrons that are spatially separated but spin-correlated. This collective spin state can be either a singlet or triplet state, which differ in the alignment between the two spins. The initial spin state is then modulated through the interaction of the electron spins with their environment. There are a number of different interactions that can take place to drive this modulation. The radical pair mechanism, as it appears in the context of quantum biology, often prioritises two of these interactions: the Zeeman and hyperfine effects. The Zeeman effect is the interaction of electron spin with a static external field, such as the geomagnetic field. The hyperfine effect is the interaction of electron spin with the magnetic moments of the surrounding nuclei in nearby amino acids, such as tryptophan or cysteine. These interactions result in the radical pair oscillating between singlet and triplet states. The final step of the radical pair mechanism is the conversion of this spin character into a measurable chemical outcome. The radical pair is an intermediate state. Chemical binding is restricted by the spin state, which means that a magnetic field can change the outcome of a chemical reaction by changing the spin state of the intermediate radical pair \cite{steiner,horerodgers,horespin}. Thus the ratio of singlet to triplet states, determined by the Zeeman and hyperfine interactions, can alter the yields of chemical reactions. This means that changing the parameters of the external magnetic field (travel to a new planet) or the local hyperfine fields (different molecules in the radical pair vicinity) will change the ratio of singlet to triplet, in turn changing the chemical outcome and any biological function dependent on this outcome. Travel to any of the other planets in our solar system will result in some perturbation to this singlet to triplet ratio, as the magnetic fields of the different planets range from negligible to stronger than the Earth's field.}
	\label{fig:RP}
\end{figure}
\\
Not long after the initial discoveries and development of spin chemistry \cite{steiner,horespin}, the radical pair mechanism was adopted as a potential mechanism to explain the avian compass \cite{schulten78}. The avian compass appears to be unaffected by a switch in North and South poles, which has led to the suggestion that it is an inclination compass, and location is determined by the angle of the Earth's magnetic field lines \cite{horerodgers}. The radical pair mechanism is currently the model that best describes this, and other, aspects of the avian compass -- although there has been some contention that a three radical model would address some of the many issues that plague the two radical model \cite{kattnig1,kattnig2}. In short, it is hypothesised that light incident on cryptochrome proteins in the eyes of birds cause the excitation and transfer of an electron, forming a radical pair. The alignment of the bird in the Earth's magnetic field then modulates the ratio of singlets and triplets through the Zeeman effect, which ultimately reflects in a different chemical outcome that translates into a specific biological signalling state \cite{schulten78,mouritsen2005,ritz2000,horerodgers}. There have been many papers published on the topic, though there remains an absence of definitive experimental verification, with the majority of experiments targeting bird behaviour rather than chemical or biological markers \cite{horerodgers,ritzres,ritzhore}. Although a recent study demonstrated that flavin-based autofluorescence in cells is sensitive to weak magnetic fields, a hallmark of the radical pair mechanism \cite{ikeya}. The main candidate molecule for radical pair chemistry in biological systems has been cryptochrome, in which the radicals formed involve flavin adenine dinucleotide (FAD), tryptophan residues and potentially superoxide \cite{simon,ritz2000,ritz2009,solovyov1}. While most of the focus has remained on cryptochrome, there are a number of other potential candidates for radical production involving amino acid residues such as tyrosine \cite{simon}. For an excellent review of radical pairs in biology, including potential radical candidates, see the paper by Zadeh-Haghighi and Simon \cite{simon}. The radical pair model of avian magnetoreception has been a subject of study for over four decades. However, in the past few years there has been growing interest in applying the radical pair mechanism to novel biological contexts \cite{simon1,simon2,simon3,simon4} such as modulation of ROS.

\subsection{Reactive oxygen species (ROS)}
It is perhaps surprising that it has taken so long for interest in the radical pair mechanism to expand into new contexts, particularly in the context of medicine, given the central role of redox chemistry in living systems. Free radicals, for instance, are commonly referred to in the physiological and pharmacological context \cite{lobo}. While they are most often referred to as having a negative, inflammatory influence on the body, these highly reactive molecules, often of oxygen species, are also important signalling molecules and part of a complex homeostatic system \cite{sies}. Certain of the models employed in the avian compass context have speculated that the radical pair mechanism may involve an oxygen radical \cite{ritz2009,solovyov1}. But there has been very little research done on how magnetic fields might modulate the production and balance of ROS in a physiological context. Recent interest in the subject, however, has given rise to a number of studies. ROS are inextricably coupled to metabolic concerns, with mitochondrial electron transport chains being a primary source \cite{sies,usselman2}. Indeed a recent study investigating magnetic effects in recombinant human electron transfer flavoenzyme reoxidation shows that static magnetic fields modulate ROS partitioning, with changes to hydrogen peroxide and superoxide production. The authors of the study suggest that this points to a flavoprotein based magnetic sensor that is linked to mitochondrial bioenergetics and cell physiology \cite{usselman,usselman2}.\\
\\
The radical pair has been suggested to play a role in a variety of new biological contexts ranging from the mechanism of lithium action in bipolar disorder to anaesthesia, to microtubule reorganisation and circadian rhythms \cite{simon,simon1,simon2,simon3,simon4}. In particular, the radical pair model has also been invoked to explain the effects of ROS on adult neurogenesis and cognition in mice \cite{rishabh}. Experimentally, this hypomagnetic field is achieved by attenuation of the geomagnetic field \cite{zhang1,zhang2}. Use of a two radical model, however, has elicited some criticism for the fact that the radical in question would be created in a triplet rather than a singlet state. In addition to this, lifetimes of the oxygen radical in question are very small, which throw some doubt on the efficacy of the mechanism \cite{kattnig}. A radical triad would mitigate some of these effects, and is suggested to play a role in hypomagnetic field effects on neurogenesis \cite{kattnig}.\\
\\
What is interesting is that, in addition to neurogenesis, a number of the studies investigate the effects of weak magnetic fields on stem cells in other contexts. One study reviewed evidence for a link between non-ionising radiation and apoptosis-induced cell proliferation \cite{beane}. Another investigated whether static weak magnetic fields might change stem cell proliferation and differentiation, through modulation of ROS and heat shock proteins. The results demonstrated different field strengths increased or decreased tissue formation in vivo \cite{vanhuizen} and furthermore that this specifically involved modulation of superoxide species \cite{kinsey}. Superoxide has also been shown to play a pivotal role in mammalian magnetic field signal transduction relating to circadian rhythms \cite{egg}. As reported in a recent review, magnetic fields could play an important role in stem-cell based regenerative therapeutics as well as controlled drug delivery \cite{marycz}. In addition to these motivating studies, the correct balance of ROS is vital in the effective functioning of biological systems. As such, a better understanding of how magnetic fields interact with these molecules seems essential.\\
\\
Quantum biology research has mostly investigated magnetic mechanisms in their overtly functional context, such as magnetoreception. What is less well documented is how the Earth's magnetic field is written into the materials out of which biology is assembled, through the interaction between the Zeeman effect (external magnetic field) and the hyperfine interaction (magnetic properties of the material). The radical pair mechanism responds not only to external fields, but also to the magnetic moments of surrounding nuclei and thus to the molecular environment in which it is found. This has led to studies investigating how changing this molecular environment might change radical pair dynamics and subsequent chemical outcomes \cite{simon,simon1,simon2,simon3,simon4}. These changes have mostly taken the form of isotope substitution, where the changed nuclear spin modifies the hyperfine interaction. In this paper our interest is in the effects that different magnetic field environments -- space travel for example -- will have on radical pair yields, and as a corollary to this, how the specific geomagnetic field shaped the evolution of life on Earth. For this reason we investigate parameter regimes for both Zeeman and hyperfine interactions, in an attempt to gain some insight into how the endogenous properties of biological materials -- the hyperfine coupling strengths -- might be tuned to specific exogenous fields. This might also offer us some new insight into magnetic effects on Earth, where the strength of the magnetic field varies by latitude, and how Earth-bound biological systems have evolved some robustness against these variations.

\section{Part one: A simple example}
\subsection{Theory}
In this section our main aim is to demonstrate the effects of different external magnetic fields, such as those experienced in space, on biological organisms. In the context of physiological health we are less interested in magnetoreception than we are in radical reactions and ROS, due to their role in inflammatory and signalling processes. For this reason we focus on measures of singlet and triplet states, rather than magnetic sensitivity. Spin state dictates the reaction products of specific radical reactions. For example, in a recent paper investigating the role of hypomagnetic fields on neurogenesis, the authors employ a flavin adenine dinucleotide (FADH) and superoxide (O2-) radical pair in which the singlet state gives hydrogen peroxide, whereas the triplet state results in superoxide product \cite{usselman,rishabh}. We thus derive an expression for spin state probability in terms of hyperfine and Zeeman parameters. To keep this tractable we start with a very simple radical pair model, in which we retain only the hyperfine interaction and the Zeeman interaction. The full spin Hamiltonian for a radical pair includes other terms and is given by \cite{tiersch}:
\begin{equation}\label{spinham}
H_S=H_{\mathrm{Zee}}+H_\mathrm{hf}+H_\mathrm{dip}+H_\mathrm{ex}+H_\mathrm{nuc},
\end{equation}
where $H_\mathrm{Zee}$ and $H_\mathrm{hf}$ are the Zeeman and hyperfine effects. We neglect to include $H_\mathrm{nuc}$, the interaction of the nuclear spins with the external field, as the gyromagnetic ratio is much smaller than in the electronic case \cite{tiersch}. In many of the models employed in radical pair research, the dipole and exchange effects, $H_\mathrm{dip}$ and $H_\mathrm{ex}$, are commonly discounted either due to sufficient separation of the electrons or because the separation is optimal for the effects to cancel \cite{horeexchange,solovyov}.  However, one of the relevant criticisms of quantum biology is the sometimes overly simplistic approach of the models used, especially given the complex, multifaceted nature of biological systems. Indeed, it has even been suggested that the conventional radical pair is unlikely to satisfy the many conditions required to be an effective sensing mechanism \cite{kim}. To this end, a three radical model has been proposed to be a more efficient iteration of the mechanism \cite{kattnig1, kattnig2}. With this in mind we acknowledge that the model we employ in this paper is a toy model. This model, however, has been used in a number of other papers investigating the radical pair mechanism in a biological context and serves to at least demonstrate the effects of modulating the external magnetic field.\\
\\
The model entails a pair of electrons that are spatially separated but spin correlated, being formed in a singlet state. The simplified Hamiltonian for the two electrons, labelled (1) and (2), each interacting separately with their nuclear environment can be written as
\begin{equation}\label{hamnc}
H_s = \gamma_e(\vec{B}.\textbf{S}^{(1)}+\vec{B}.\textbf{S}^{(2)})+
\sum_{n=1}^2\sum_{i=x,y,z}A^{(n)}_i\textbf{S}_i^{(n)}\otimes\textbf{I}_i^{(n)},
\end{equation}
where $\hslash$ is taken to be 1 and $\vec{B}$ is the magnetic field vector. The first term represents the Zeeman interaction with electron gyromagnetic ratio $\gamma_e =-g\mu_B$, where $g=2$ and $\mu_B$ is the Bohr magneton. The second term gives the hyperfine interaction, where each electron interacts with its own nuclear environment and $A_i$ represents the elements of the hyperfine coupling tensor. Once again $\textbf{S}=(S_x,S_y,S_z)$ is the vector of spin operators for each electron with  $S_i=\frac{1}{2}\sigma_i$, $i=x,y,z$, where $\sigma_i$ are the Pauli matrices and $\textbf{I}=(I_x,I_y,I_z)$ is the vector of spin operators for nuclear spin. For the majority of the results in this section we assume that one of the electrons in the pair does not interact with any nuclei. In this sense we follow the common approach adopted in other papers \cite{gauger} for simplicity of calculation. However, we also do so because we are interested in the spin dynamics of reactive oxygen species, where only one of the radicals in the pair experiences the hyperfine interaction. Much of the radical pair literature in quantum biology focuses on radical pairs in cryptochrome proteins, in the context of the avian compass. These most often involve flavin and tryptophan radicals. However, some research suggests that oxygen radicals may also play a role \cite{ritz2009,solovyov1}. For singlet product yield we follow the example of Tiersch and Briegel \cite{tiersch}
\begin{equation}
\Phi_S=\int_{0}^{\infty}dt k e^{-kt}\braket{S| \rho_e(t)| S},
\end{equation}
where $S$ is the singlet probability, $\rho_e(t)$ is the reduced density matrix of the electrons and $k$ is the radical pair lifetime. For the purposes of this paper, which is to illustrate the basic concept of the radical pair response to external fields, we do not include noise.

\begin{figure}
        \center
	\includegraphics[scale=0.65]{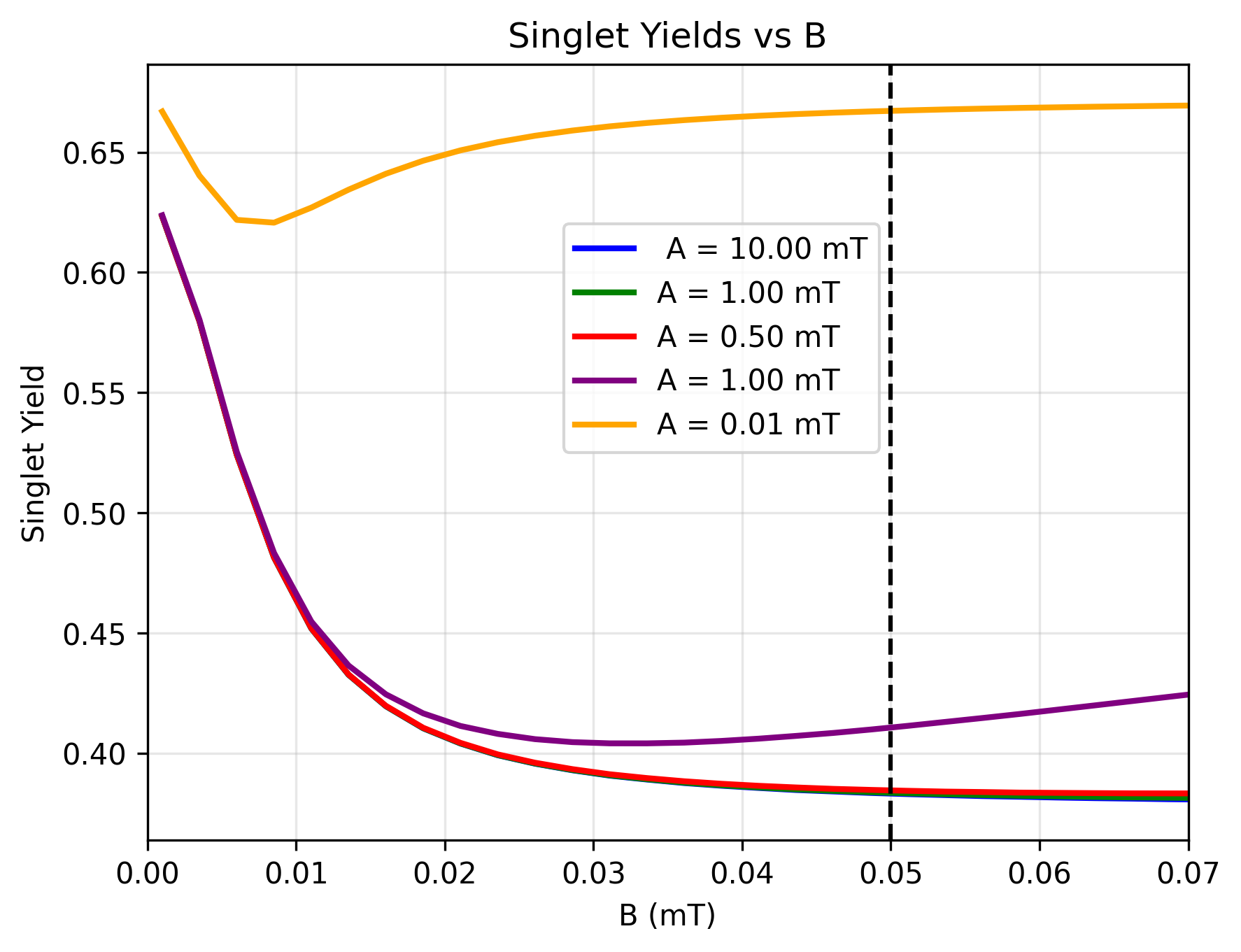}
	\caption{Graph showing the dependence of singlet product on an Earth strength magnetic field, for a range of isotropic hyperfine coupling strengths $A_x=A_y=A_z=A$. For smaller hyperfine coupling constants the singlet yield reaches a minimum for smaller magnetic fields before increasing as the field strength approaches Earth strength (given by the vertical dashed line). Increasing the hyperfine coupling constants, in turn increases the field strength at which this minimum occurs, as well as decreasing absolute singlet yield. It is also clear that singlet yield varies across the geomagnetic range of about 20 to 70 $\mu$T, potentially offering an explanation as to why some diseases manifest differently at different latitudes. The graph also gives us some idea of the effects of hypomagnetic, close-to-zero magnetic fields, such as those on the Moon or Mars. Even without the Zeeman effect, which acts to lift the degeneracy of the triplet states, singlet-triplet mixing occurs due to the hyperfine interaction \cite{simon5}. We have based our range of hyperfine coupling constants on the information for organic molecules in the paper by Hiscock \textit{et al}. \cite{hiscock}}
	\label{fig:Earth}
\end{figure}
\begin{figure}
        \center
	\includegraphics[scale=0.65]{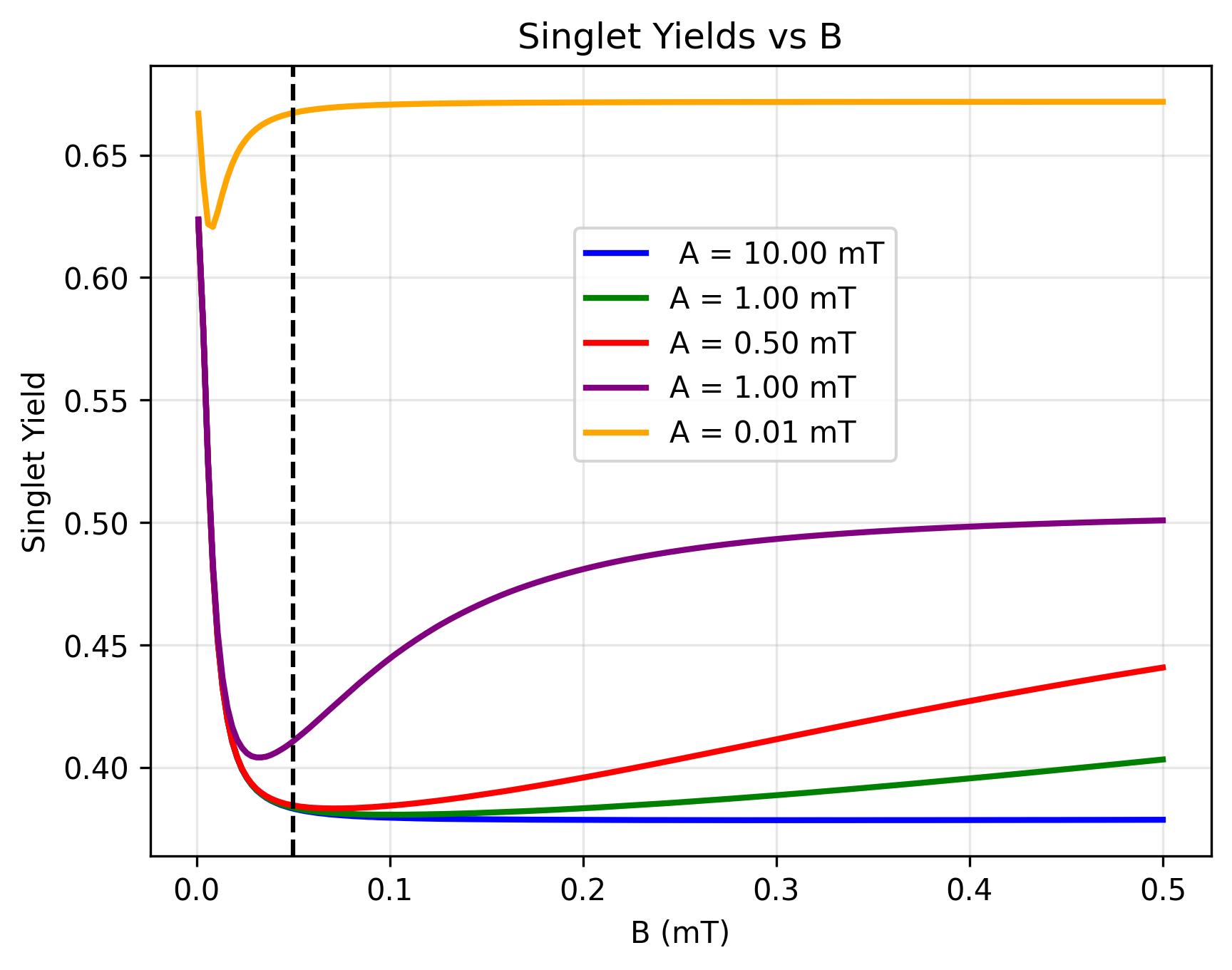}
	\caption{Graph showing the dependence of singlet product of the magnetic fields of other planets. To further illustrate the relationship between reaction yield, magnetic field strength and isotropic hyperfine coupling $A_x=A_y=A_z=A$, we plot singlet yield across magnetic field strengths corresponding to those planets in our solar system, which range from zero for planets such as Mars to the strong magnetic field of Jupiter, in comparison to Earth strength (given by the vertical dashed line). We investigate hyperfine coupling constants within the range of organic materials but also coupling strengths an order of magnitude larger. Singlet yield depends on the ratio of hyperfine coupling constant to magnetic field strength, with smaller hyperfine couplings giving bigger singlet yields. For larger magnetic fields the singlet yield reaches a minimum at larger hyperfine coupling strengths.}
	\label{fig:Iso}
\end{figure}
\begin{figure}
        \center
	\includegraphics[scale=0.65]{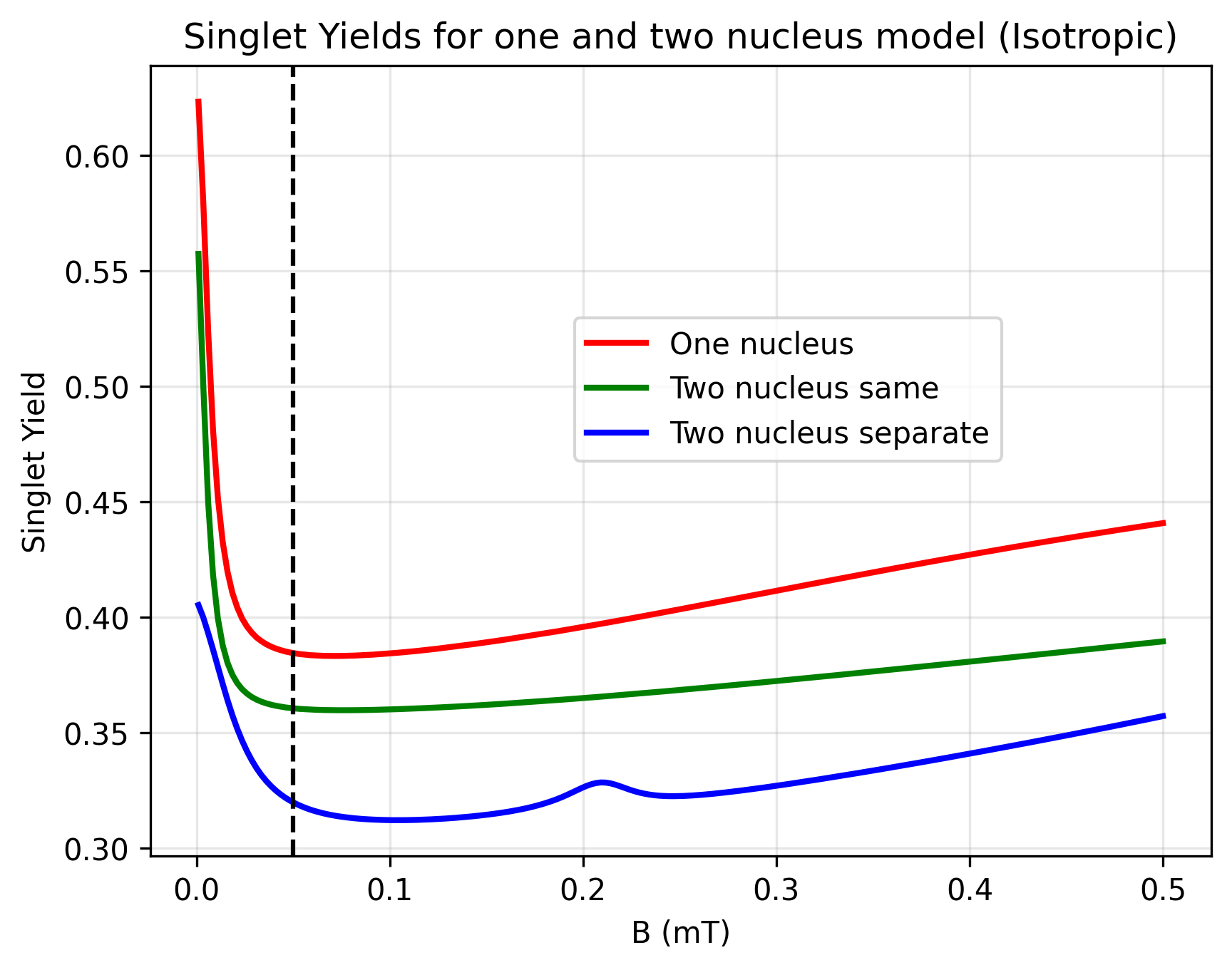}
	\caption{Graph showing the dependence of singlet product on different hyperfine spin environments. To investigate the case for more complex hyperfine environments we include an additional nucleus, with spin 1 rather than $\frac{1}{2}$, as compared to a single spin $\frac{1}{2}$ nucleus interacting with only one of the radicals. We consider two cases: two nuclei on one of the radicals (green graph) and one nucleus on each of the radicals (blue graph), over a range of external magnetic fields, with Earth's field given by the vertical dashed line. In the former case there is a smaller overall singlet yield than the single nucleus case but this reaches a minimum at a magnetic field strength less than 100 $\mu$T, in a manner similar to the single nucleus case. For the case in which each radical in the pair interacts with a nucleus, the singlet yield is further decreased. In addition to this the yield has a small peak at 200 $\mu$T rather than a steady increase as in the other two cases. We are predominantly interested in the green graph given that we are considering one of the radicals to be an oxygen radical, without hyperfine interaction.}
	\label{fig:Iso2}
\end{figure}
\subsection{Results}
There has been growing interest in using the radical pair mechanism to model diverse biological functions \cite{simon,simon1,simon2,simon3,simon4}. In most of these instances the focus of interest has been the role of the hyperfine effect and how this changes, for instance, due to isotope substitution. More recently, the radical pair mechanism has been used to model the effects of a hypomagnetic field, by looking more closely at how the Zeeman effect modulates spin dynamics \cite{simon5}. We are interested in the outcome of radical pair dynamics under different external magnetic fields, given the possibility of space travel in hypomagnetic fields as well as settlement on planets with diverse magnetic fields. Although there is some contention as to the lifetime of radical pairs in biological systems, we chose a lifetime of 10 $\mu$s for the following reasons. Gauger \emph{et al}. estimate a lifetime of 100 $\mu$s to allow sufficient time for appropriate spin state mixing \cite{gauger}. A number of other papers put the lifetime a few orders of magnitude smaller. We have chosen a lifetime between these two extremes. We first investigated radical pair dynamics for the Earth's magnetic field which ranges from around 20 to 70 $\mu$T. As an output we plot singlet yield for the different magnetic field strengths to illustrate to what extent the initial singlet state has been converted into a triplet state. Although the potential of intermediate triplet states to result in ROS has been categorised as harmful to biological systems, ROS play a more complex role in biology, being important signalling molecules \cite{sies}. This flexibility of purpose could be partially facilitated by the variability of the endogenous parameters of organic radicals in an Earth strength field. To put it more simply: different radical pair candidates have different coupling strengths, which results in different singlet to triplet ratios and thus a different balance of ROS species and concurrent biological functionality. Hyperfine coupling constants for organic molecules range from around 10 to 1000 $\mu$T. We have plotted singlet yields over this range.\\
\\
For smaller hyperfine coupling constants the singlet yield reaches a minimum for smaller magnetic fields before increasing as the field strength approaches Earth strength. For coupling constants of this strength the yield is predominantly singlet. Increasing the hyperfine coupling constants, in turn increases the field strength at which singlet yield reaches a minimum. Maximum singlet-triplet conversion appears to occur mostly within the bounds of organic parameters and Earth-strength fields. It is also clear that, at least for the toy model employed in this paper, singlet yield varies across the variable field of the Earth, potentially offering an explanation as to why some diseases manifest differently at different latitudes. See Figure 2 for details and Section 4.2 for further discussion of latitude-dependent diseases. To further illustrate the relationship between magnetic field strength and hyperfine coupling, we plotted singlet yield across magnetic field strengths corresponding to those planets in our solar system. For the most part, we kept the hyperfine coupling constants within the range of organic materials but also investigated coupling strengths an order of magnitude larger. In figure 3 it can be seen that for larger magnetic fields the singlet yield reaches a minimum at larger hyperfine coupling strengths. In this sense singlet-triplet mixing in biological materials that evolved on Earth reflect the interplay between the Earth's field and the magnetic properties of these materials. This does not mean that other versions of biology would not have evolved in a different magnetic field, but only that the magnetic parameters of biological materials as we know them, were optimised in relation to this specific field. To see whether this persists for more complex hyperfine environments we investigated the inclusion of an additional nucleus, with spin 1 rather than $\frac{1}{2}$, as compared to a single spin $\frac{1}{2}$ nucleus interacting with only one of the radicals. Figure 4 shows results for the case of two nuclei on one of the radicals as well as one nucleus on each of the radicals. In the former case the singlet yield is decreased but reaches a minimum at a magnetic field strength less than 100 $\mu$T, in a manner similar to the single nucleus case. For the case in which each radical in the pair interacts with a nucleus, the singlet yield is further decreased. In addition to this the yield has a small peak at 200 $\mu$T rather than a steady increase as in the other two cases. While this may have relevance for other biological contexts in which radical pairs feature, for the purposes of this paper we are particularly interested in the case of oxygen, where one radical experiences no hyperfine environment. 

\section{Part two: terrestrial fine-tuning}
\subsection{Theory}
In the first part of this paper we demonstrated the effects of different external magnetic fields on the yield of ROS species for a simple radical pair model --interacting with one, then two, nuclei -- with isotropic hyperfine interactions. In this second part of the paper we look in more detail at how the intrinsic magnetic parameters of biological materials, given by their hyperfine coupling strengths, are tuned to the environment these materials evolved in, such as the Earth's magnetic field. To do this we look at how the spin state probabilities depend on the interplay between the parameters of the spin Hamiltonian. For the simple Hamiltonian given by Equation 2, these parameters are the Zeeman and hyperfine terms. The Zeeman parameter is determined by the strength of the external field which is independent of the material properties of living organisms. The hyperfine parameter, on the other hand, depends on these intrinsic material properties, as represented by the hyperfine tensor $A^{(n)}_i$ in Equation 2. The hyperfine tensor includes the hyperfine coupling constant, which represents the strength of the interaction, and the anisotropy which is given by the $A_x$, $A_y$, and $A_z$ components of the hyperfine tensor. The exact values of the tensor depend on the specific nuclear environment of the radical pair, such as the positions of the interacting nuclei, or the type of nuclei, for instance hydrogen and nitrogen will have different coupling strengths. The results in the first part of this paper demonstrate that radical pair yields depend on the strength of the hyperfine coupling, with respect to the strength of the external field. For example, coupling strengths much bigger than the external magnetic field have very different dynamics to those of comparable strength, see Figures 2 and 3 for details. While Figures 2--4 show results for isotropic coupling, the radical dynamics also depend on the anisotropy of the coupling, which we now consider. We are particularly interested in how this anisotropy can optimise certain biological outcomes. In a paper on avian magnetoreception, for example, Hiscock \textit{et al} discuss a sharp spike in compass sensitivity that depends on the anisotropy of the hyperfine interaction, specifically the large $A_z$ component and smaller $A_x$ and $A_y$ components. They attribute this to avoided crossings, which result from the small perturbation of the $A_x$ and $A_y$ components that couple the diagonal states of the spin Hamiltonian, for details see the supplementary material for the paper by Hiscock \textit{et al} \cite{hiscock}. In this paper we are interested in the ways in which the anisotropic elements of the hyperfine tensor interact with the external magnetic field to optimise singlet or triplet product in the context of ROS balance, rather than compass sensitivity. To get a better idea of exactly how hyperfine and Zeeman parameters interact, we derive a simple expression for singlet state probability as a function of these parameters. 
From the Liouville-von Neumann equation,
\begin{equation}
\frac{\mathrm{d}}{\mathrm{dt}}\rho(t) = -\frac{i}{\hslash}[H_S,\rho(t)],
\end{equation}
$H_S$ is the system Hamiltonian given in Equation 2 and $\rho$ is the density operator, we solve for the time dependent density matrix. The density operator evolves as 
\begin{equation} 
\rho(t)=U\rho(t_0)U^{\dagger}.
\end{equation}
where $U(t)$ is the time evolution operator:
\[
U(t) = e^{-itH}.
\]
The density matrix tracks the evolution of all four possible spin states -- one singlet, three triplets -- of the radical pair. To look specifically at the singlet state dynamics we use the fact that singlet probability is given by $\langle{S}\rangle=\mathrm{Tr}\{S\rho\}$. For the simplest case of a magnetic field in the $z$ direction and anisotropic coupling, $A_x = A_y=0$, the singlet state time evolution is given by,
\begin{equation}
S(t) = \frac{1}{4}(3+\mathrm{cos}(4A_zt)).
\end{equation}
As is to be expected, for the case in which only the $A_z$ component of the hyperfine tensor is considered, the spin dynamics are independent of an external magnetic field in the $z$ direction, with Zeeman term $\gamma_e(B.S_z^{(1)}+B.S_z^{(2)})$ for both electrons of the radical pair. In the slightly more complex case in which $A_x=A_y=A$ and $A_z=0$, in the same external field, the singlet dynamics are dependent on both Zeeman and hyperfine parameters,
\begin{equation}
S(t) = \biggl{[}1-\frac{2A^2\mathrm{sin}^2\sqrt{\gamma_e^2B^2+4A^2}t}{\gamma_e^2B^2+4A^2}\biggr{]}^2.
\end{equation}
The Zeeman term in a fixed environment such as the Earth's magnetic field, is to a large degree constant -- ignoring perturbations due to solar storms and changes over large time scales. Living organisms that have evolved within this field, have less fixed parameters: flexible to adaptation and natural selection. The crucial role that ROS play in signalling and inflammation processes in the body, suggest that ROS balance may have played an important role in evolutionary fitness. That is, the nuclear environments surrounding relevant radical pairs have been selected for to optimise singlet or triplet yields in a given external field. The materials out of which living organisms have constructed themselves are thus structurally coupled to the field they evolved in, and the introduction of a different external field will disrupt this optimisation. We illustrate this point by plotting how the different parameters of interest interact with each other.

\begin{figure*}[t!]
	\centering
        \includegraphics[scale=0.5]{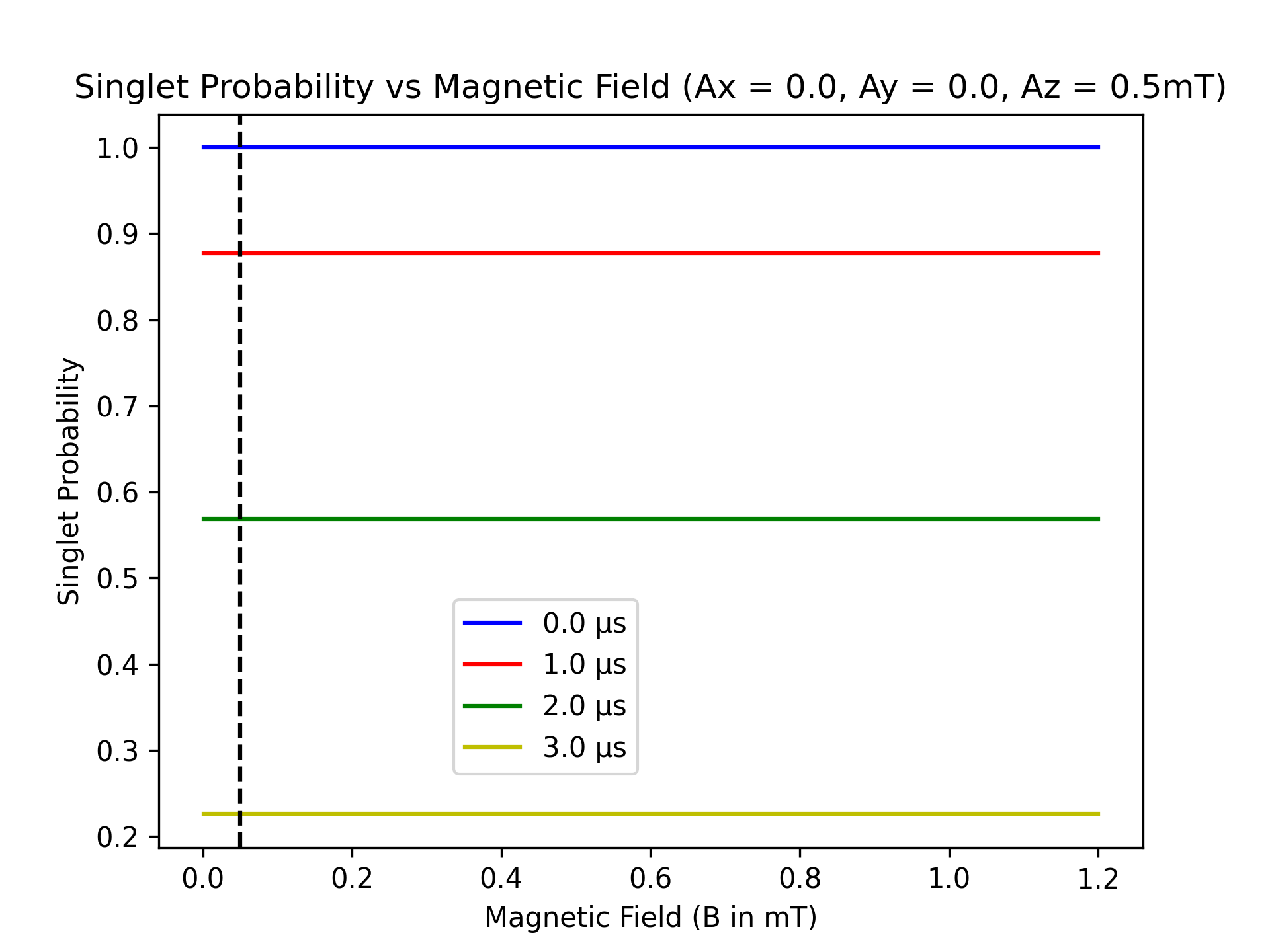}
	\includegraphics[scale=0.5]{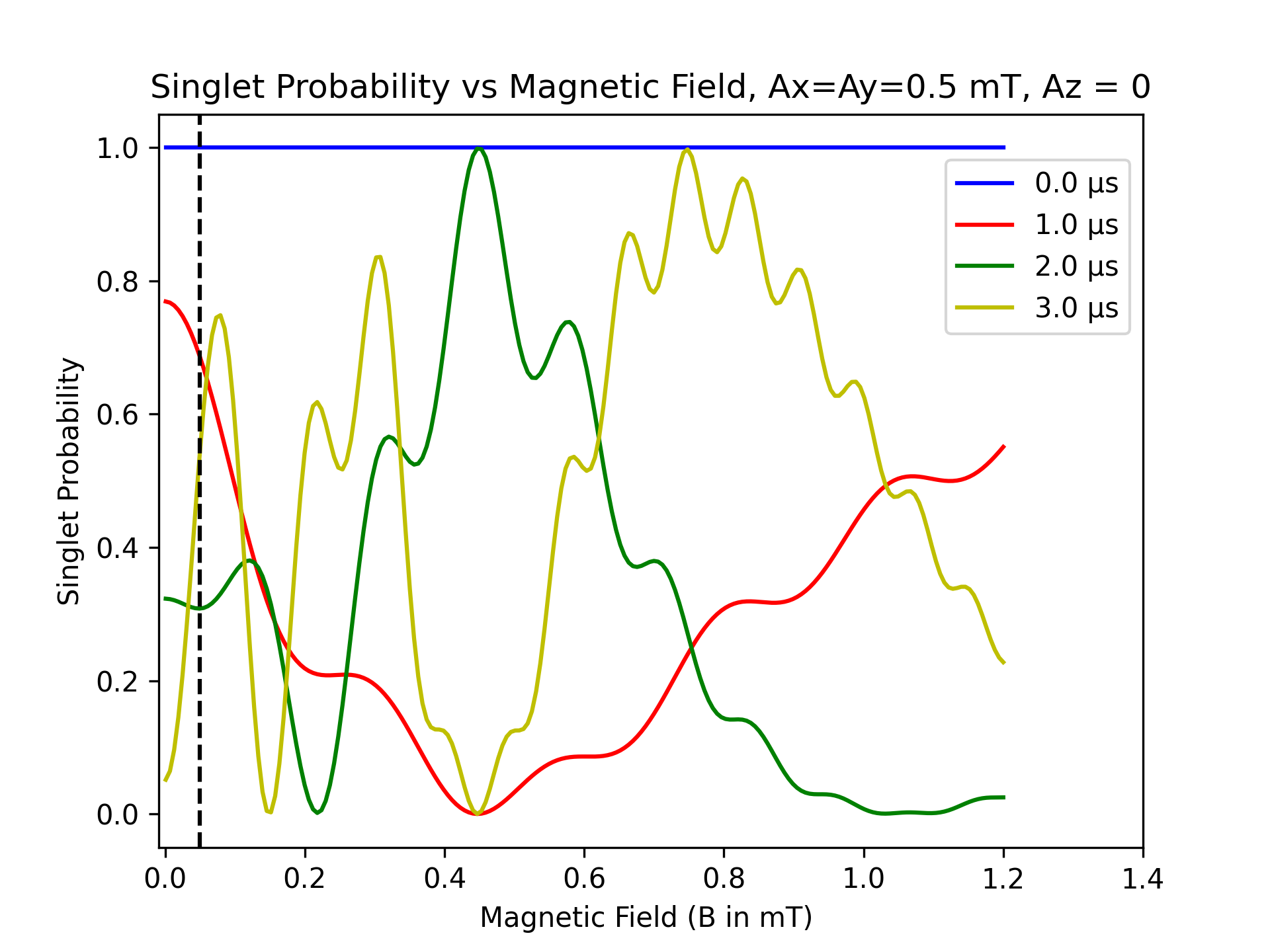}
	\caption{Graphs showing how the different components of the hyperfine tensor have very different effects on singlet probability. For the graph on the left, with $A_x=A_y=0$, the singlet is independent of an external field. This specific anisotropic configuration is interesting in the context of space travel, in that the ROS ratios would be unaffected by any novel external fields. We have taken, for comparison, four different lifetimes of the radical pair, measured in microseconds. The blue line, with $t=0$, gives the initial singlet state. For the graph on the right, setting $A_z=0$ with $A_x$ and $A_Y$ non-zero causes large fluctuations in singlet probability over the range of external magnetic field strengths, where the Earth's field is marked by the vertical grid line. It should be stated that the results should not be taken as an indication of absolute singlet probability at a specific field, given that these results represent only single points in time.}
\end{figure*}

\begin{figure*}[t!]
	\centering
        \includegraphics[scale=0.5]{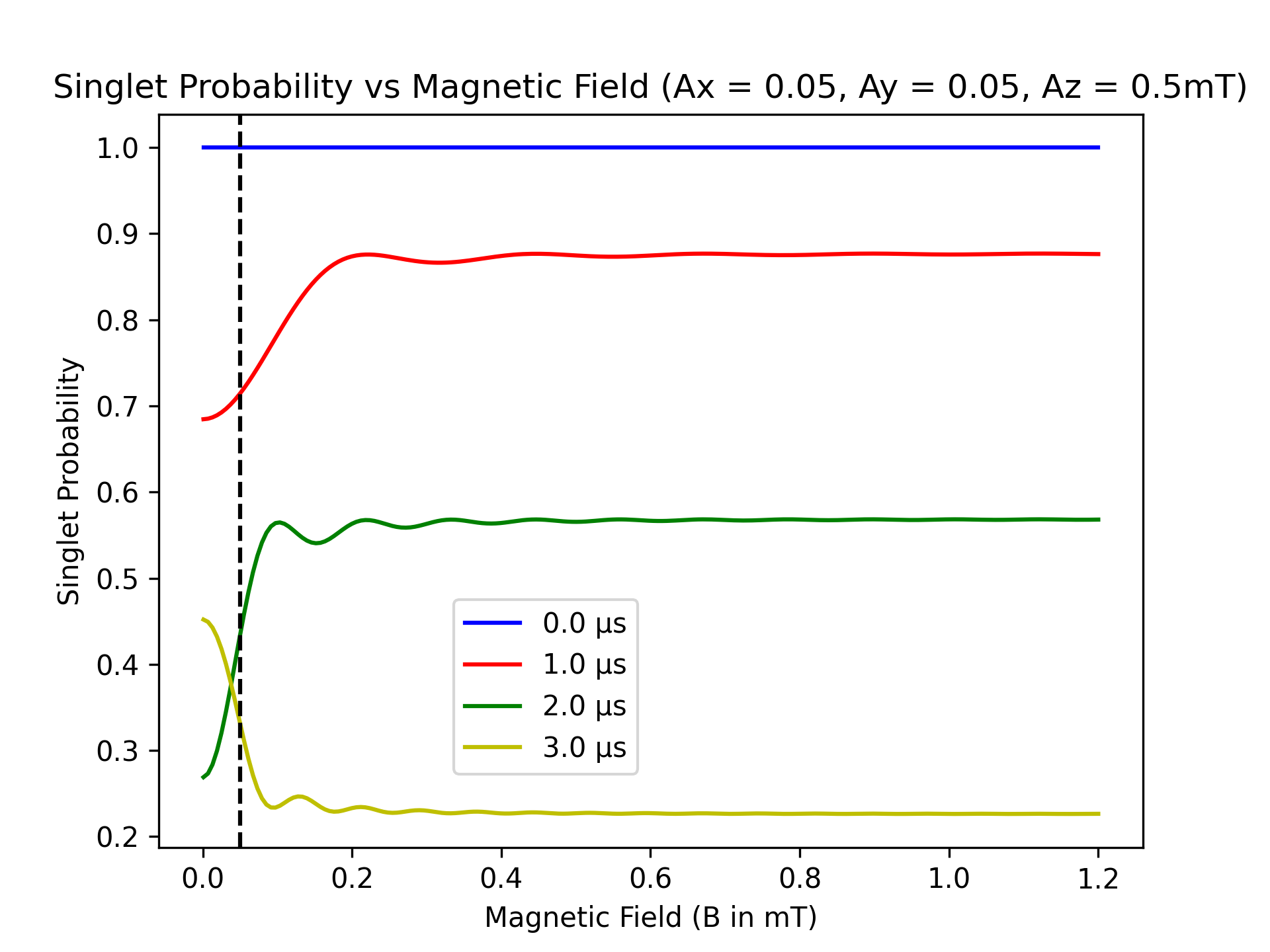}
        \includegraphics[scale=0.5]{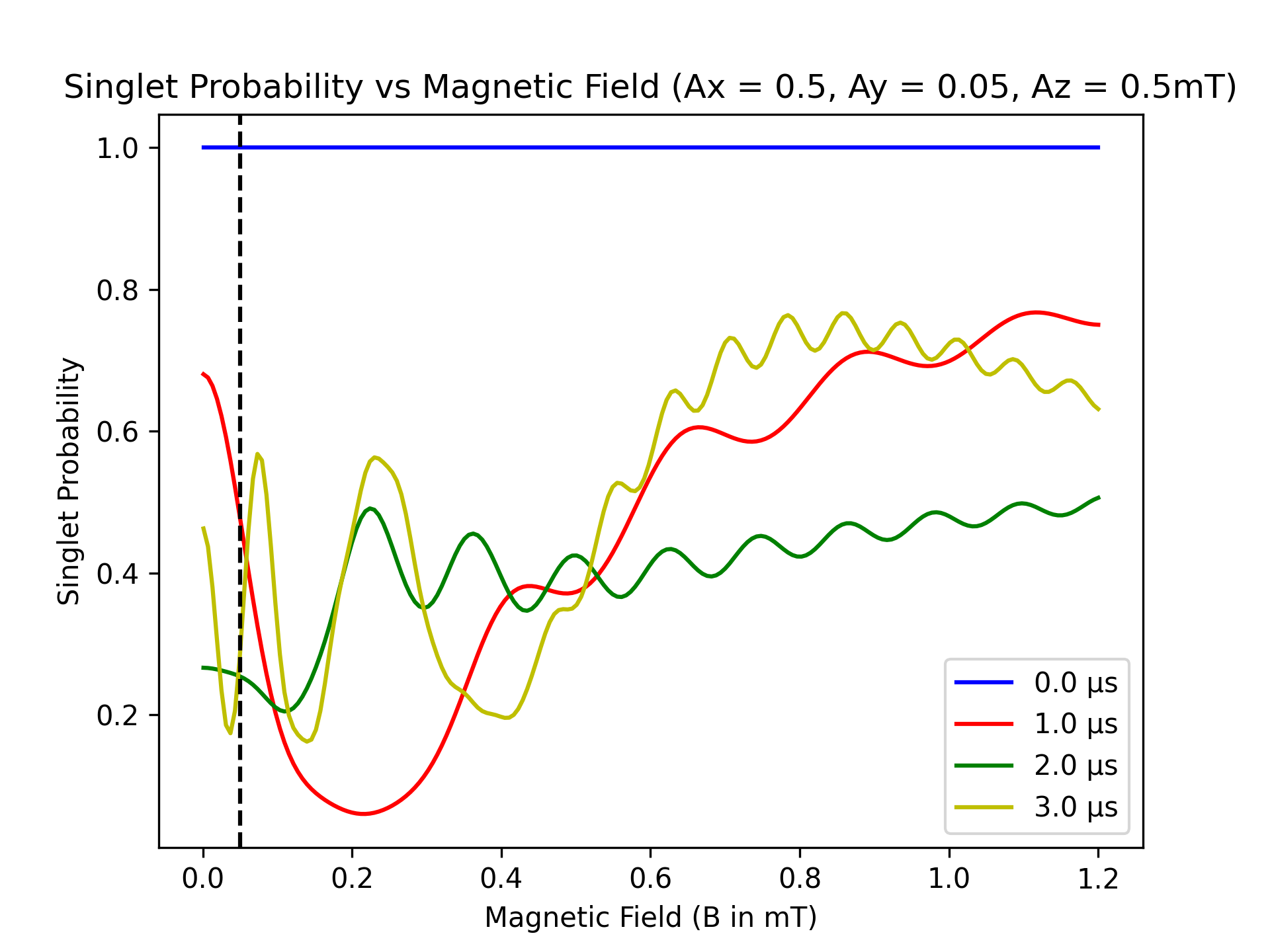}
	\caption{Graphs showing the the relationship between the different anisotropic components of the hyperfine tensor. Following observations on optimal anisotropies for compass sensitivity by Hiscock \textit{et al} \cite{hiscock} the graph on the left has small $A_x$ and $A_y$ components with a much bigger $A_z$ component. Interestingly, after an initial dependence of spin state on the external field, the graph flattens out over larger magnetic fields. This is most likely due to the fact that as the external field increases, the contribution from $A_x$ and $A_y$ begin to approximate zero, with only the $A_z$ component contributing as in Figure 5. The graph on the right illustrates the case in which one of the $A_x$, $A_y$ and $A_z$ components are much smaller than the other components. This yields a more dynamic singlet probability in relation to different magnetic fields. Again, it should be noted that these results represent single points in time.}
\end{figure*}
\begin{figure*}[t!]
	\centering
        \includegraphics[scale=0.35]{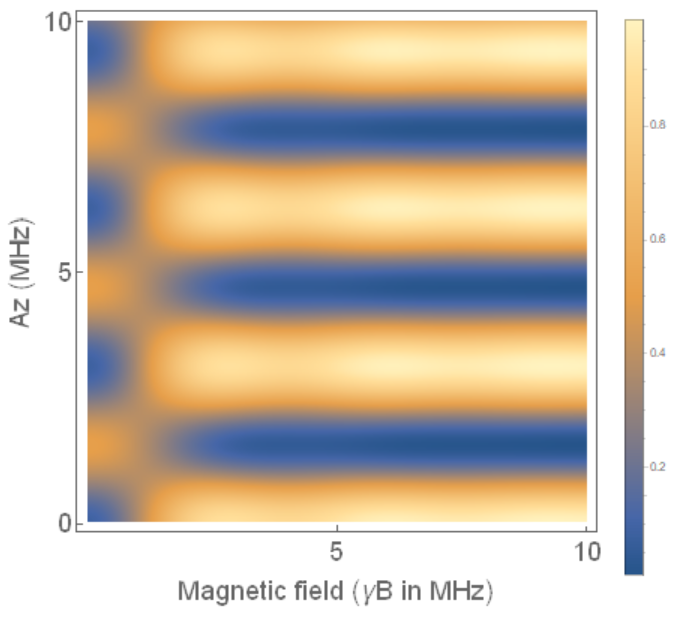}
	\includegraphics[scale=0.35]{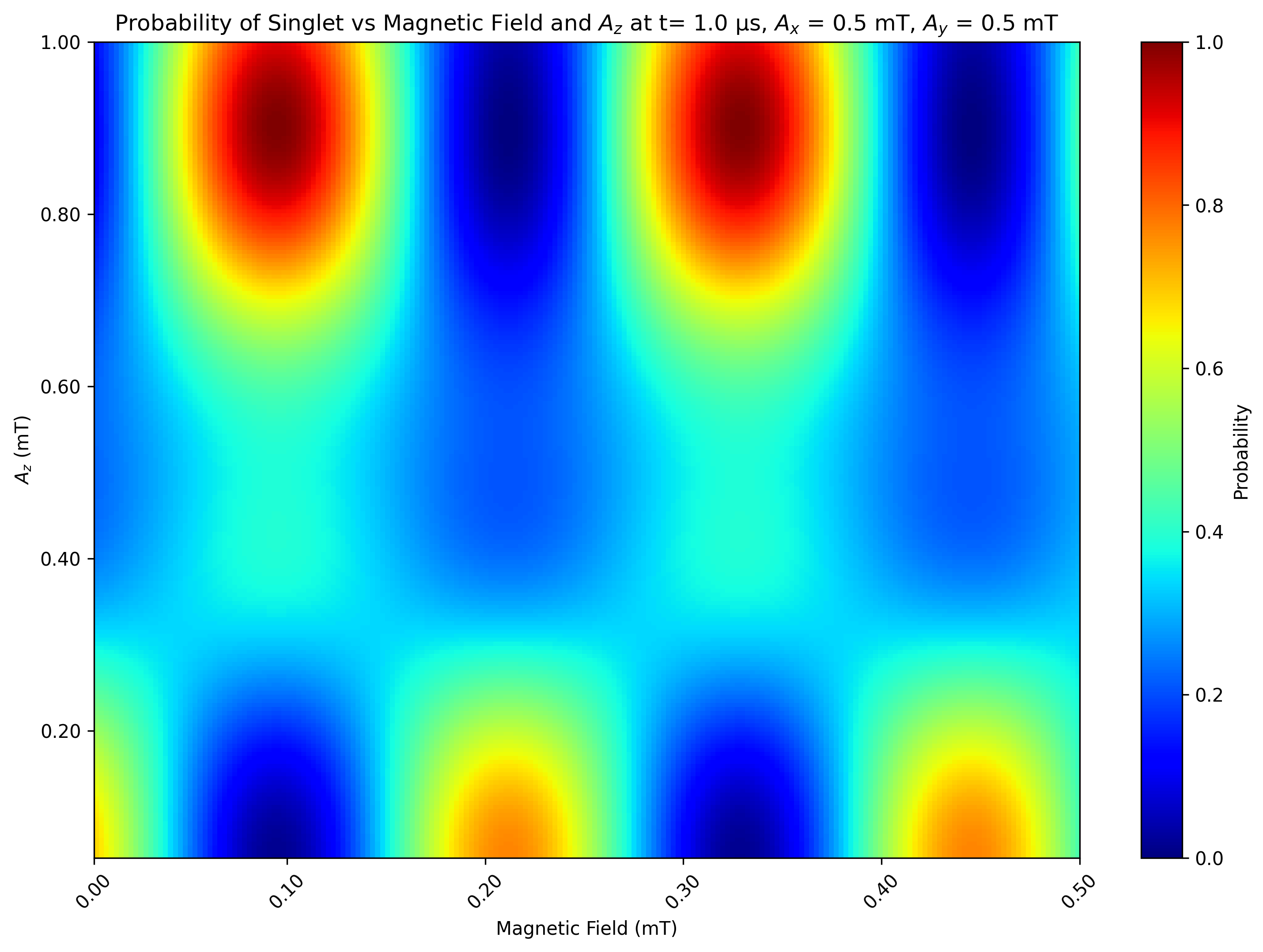}
	\caption{Graphs showing the singlet probability dependence on different external fields and $A_z$ values, for fixed $A_x$ and $A_y$. In graph on the left, for small values of $A_x$ and $A_y$ in relation to $A_z$, singlet probability for a given value of $A_z$ becomes independent of external field strength at higher field strength. This is again most likely due to the fact that as the external field increases, only the $A_z$ component plays a role, with spin probability becoming independent of the external field. In the graph on the right, for the case in which both $A_x$ and $A_y$ are large, there are distinct regions in which specific values of the $A_z$ component lead to either maximum or minimum singlet probability at different field strengths.}
\end{figure*}
\begin{figure*}[t!]
	\centering
        \includegraphics[scale=0.4]{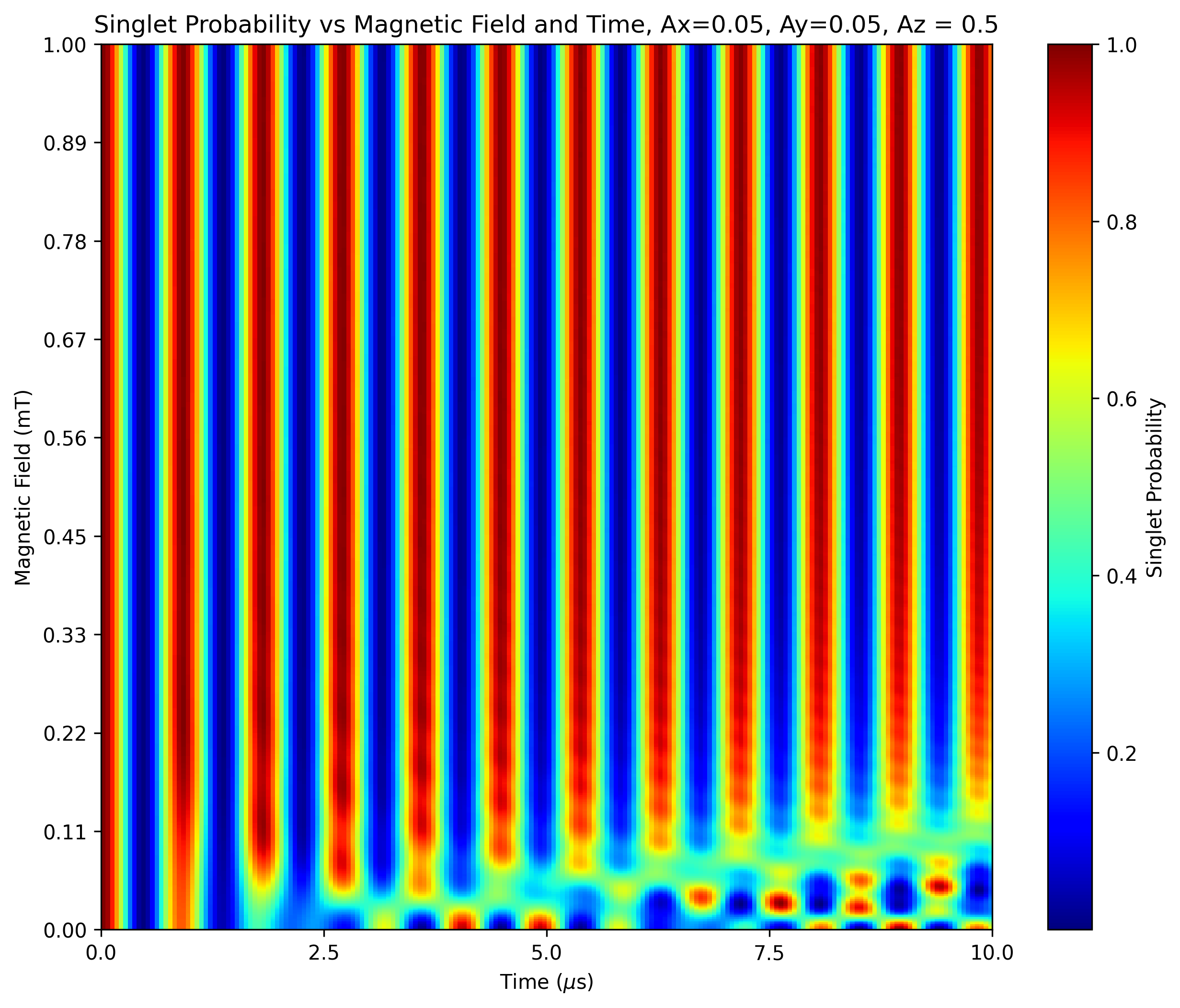}
        \includegraphics[scale=0.4]{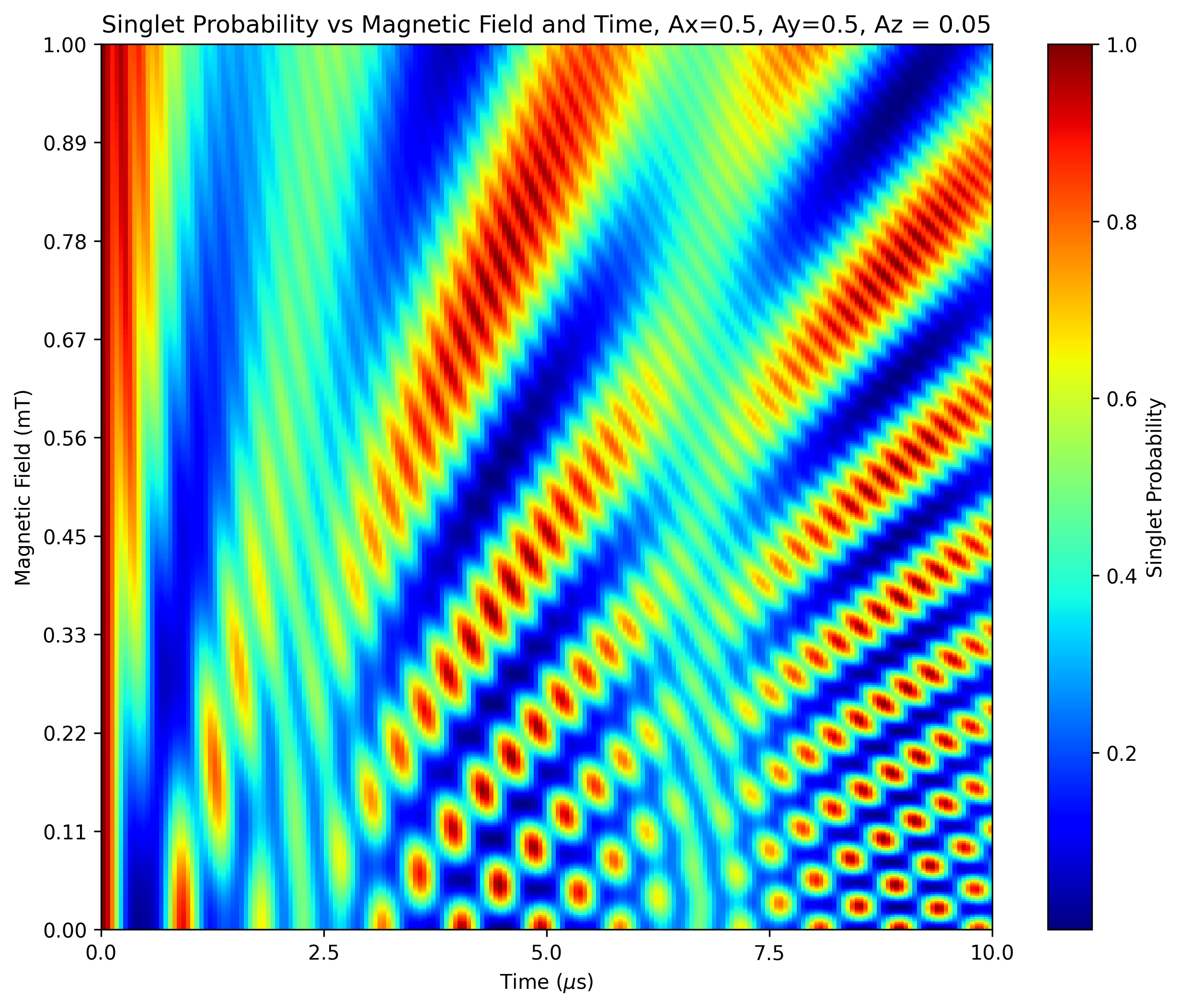}
	\caption{Graphs showing how singlet probability in different external fields evolves over a timescale of $10\mu s$, for different anisotropic combinations. Different hyperfine anisotropies  favour singlet or triplet states for different lifetimes. For example in the graph on the left, in magnetic fields of Earth strength and lower, with a small $A_x$ or $A_y$ and big $A_z$, the mixing between singlet and triplet states increases in probability over bigger timescales. However as the external field strength increases, spin state probability for a given lifetime is independent of this external field. In the graph on the right, $A_x$ and $A_y$ are big compared to a small $A_z$ (right) and singlet state for a given lifetime is strongly dependent on external field strength, compared with a small $A_x$ or $A_y$ and big $A_z$. It is also worth noting that spin state (and thus corresponding ROS balance) oscillates from maximum to minimum over increasing field strengths for a given lifetime. This would be consistent with the experimental finding that different fields can have opposite effects on ROS yields \cite{vanhuizen,beane}.}
\end{figure*}
\subsection{Results}
As a first example we plot the simple results of the expressions derived for singlet dependence on the parameters of the Hamiltonian. In Figure 5 it is clear that the different components of the hyperfine tensor have very different effects on singlet probability. For $A_x=A_y=0$ the singlet is independent of an external field. Setting $A_z=0$ with $A_x$ and $A_y$ non-zero, has the opposite effect, causing large fluctuations in singlet probability over the range of external magnetic field strengths. As a second simple example, and following from the observation that a small $A_x$ and $A_y$ component with large $A_z$, can optimise the avian compass \cite{hiscock}, we looked at the relationship between the different anisotropic components, see Figure 6 for details. It is obvious that the specific form of the hyperfine tensor is instrumental in determining the singlet probability for a given external magnetic field. We investigate this further in Figure 7, where instead of a fixed hyperfine tensor, we vary the $A_z$ component over a range of biologically relevant strengths, while fixing $A_x$ and $A_y$ at different values. The results show that for small values of $A_x$ and $A_y$ in relation to $A_z$, singlet probability for a given value of $A_z$ is close to independent of external field strength. For the case in which both $A_x$ and $A_y$ are large, there are again regions in which specific values of the $A_z$ component lead to either maximum or minimum singlet probability. These results, while potentially interesting, only show the spin probability at one point in time of $1\mu s$. We now look at how singlet probability in different external fields evolves over a timescale of $10\mu s$, for different anisotropic combinations. In Figure 8 we show that different hyperfine anisotropies  favour singlet or triplet states over time. For example, in magnetic fields of Earth strength and lower, with a small $A_x$ or $A_y$ and big $A_z$, mixing between singlet and triplet states increases in probability over bigger timescales. In the case where $A_x$ and $A_y$ are big compared to a small $A_z$, singlet state for a given lifetime is strongly dependent on the external field strength. It is also potentially interesting to consider recent experimental results that demonstrate the effects of weak magnetic fields on stem cell proliferation and differentiation, via modulation of radical spin and ROS balance. The results demonstrated different field strengths increased or decreased tissue formation in vivo \cite{vanhuizen}. Specifically, fields of roughly four times (200$\mu$T) the geomagnetic field showed decreased blastema formation in comparison to increased blastema formation in fields roughly ten times (500$\mu$T) those of the Earth's field. Figure 8 makes it clear that, at least for specific anisotropic hyperfine strengths, spin state (and thus corresponding ROS balance) oscillates over increasing field strengths of the order of those described in the stem cell experiments. This would be consistent with the experimental finding that different fields have opposite effects on ROS yields. It should be noted that we are merely interested in the principle of this oscillation, but do not assume that our graphs represent exact singlet probabilities at specific external fields, given our approximation of the hyperfine coupling. And finally, having investigated how singlet probability depends on the external magnetic field, we focus more closely on how spin state (and related ROS balance) is coupled to the Earth's magnetic field by the intrinsic magnetic parameters of living materials, such as the hyperfine coupling parameters. It is clear that singlet state is maximised or minimised for specific values of $A_z$ in relation to $A_x$ and $A_y$, as in Figure 9. In the next section we discuss how this may demonstrate quantum tuning in evolution, where materials such as amino acids, with nuclear environments that favour a certain ROS profile with specific biological outcome (signalling, stem cell moderation), might have been more likely to be selected for.

\begin{figure*}[t!]
	\centering
        \includegraphics[scale=0.6]{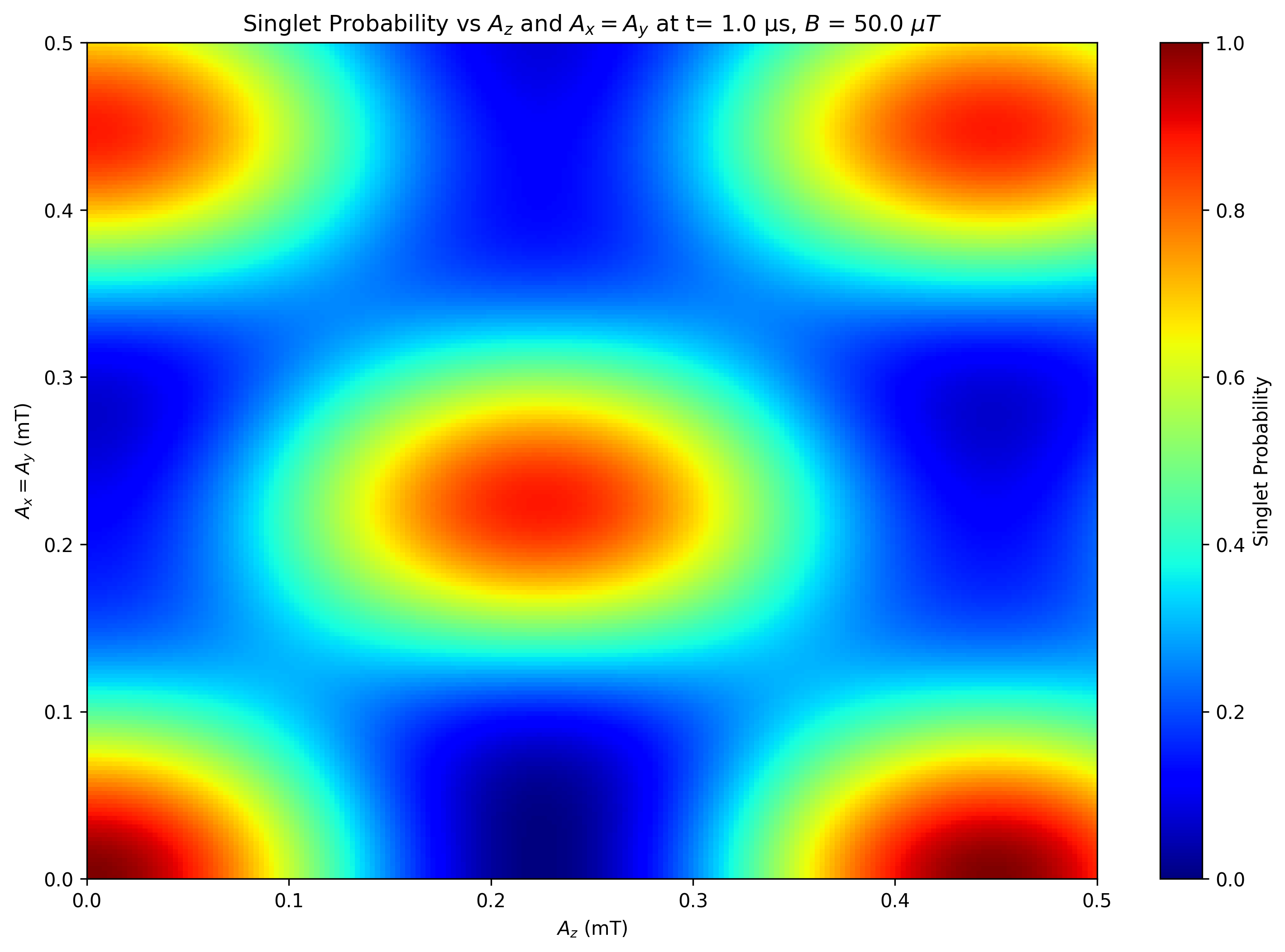}
	\caption{Graph showing how spin state (and thus related ROS balance) can be tuned to the Earth's magnetic field depending on the intrinsic magnetic parameters of living materials, such as the hyperfine coupling parameters. It is clear that singlet state is maximised or minimised for specific values of $A_z$ in relation to $A_x$ and $A_y$ within the Earth's (50$\mu$T) magnetic field. We have examined the case for hyperfine coupling strengths up to an order bigger than the Earth's field.}
\end{figure*}

\section{Discussion}
\subsection{Quantum evolution in biological materials}
There has been some interest in how quantum effects may be implicated in evolutionary processes. For example, differential metabolic fractionation of isotopes in certain species has been hypothesised to affect the evolution of migration, by improving the sensitivity of the radical pair compass \cite{galvan}. Recent interest in how avoided crossings in radical pair spin states can confer a functional biological advantage such as increased compass sensitivity in magnetoreception, also suggests that quantum effects may have played an evolutionary role in biological organisms \cite{hiscock}. The quantum needle effect investigated in the paper by Hiscock \emph{et al}., for example, suggests an evolutionary tuning of the magnetic sensitivity of the avian compass through mutation of the proteins that constitute the hyperfine environment \cite{hiscock}. They attribute this effect to two of the nitrogens interacting with the radicals. In particular they show that modulating the anisotropy of the hyperfine tensors associated with these nitrogens, in turn modulates the `needle' effect: a sharp spike in reaction yield. They suggest that this spike is a result of avoided crossings modulated by the specifics of the hyperfine tensor. In their case the sensitivity of the avian compass spikes due to the specific relative strengths of the nitrogen hyperfine interactions, with $A_x$ and $A_y$ small but non-zero in comparison to a large $A_z$. We suggest this might be a more general principle, and that evolution may have tuned the magnetic parameters of biological materials around the Earth's magnetic field, by optimising the hyperfine environment in which radical pairs operate, to exploit quantum phenomena such as avoided crossings. While the optimisation of the avian compass has drawn most of the attention, we are interested in biological function that occurs more widely across species. For this reason we have focused on how the radical pair mechanism relates to ROS balance, given the fact that ROS are important signalling molecules across all types of living organisms. We have thus chosen to focus on how, for example, the ratio of specific ROS such as hydrogen peroxide and superoxide, depend on underlying spin chemistry: the ratio of singlet to triplet states. Our simple model demonstrates how certain hyperfine environments will optimise the singlet state, while some will optimise the triplet state. This singlet-triplet interconversion is unavoidably bound to two material conditions: the strength of the external magnetic field (Zeeman effect) and the nature of the material in which the radicals occur (hyperfine effect). Hyperfine coupling constants in organic materials range in strength from smaller to larger than the Earth's field strength \cite{rodgers}. In addition to the strength of the coupling, the hyperfine interaction also varies in anisotropic parameters. For a given magnetic field (the Earth's) a specific ROS balance will be determined by the hyperfine parameters, which is in turn determined by the specific nuclei in the vicinity of the radicals. Molecules that optimise this ROS profile for organism survival, will thus be selected for. Disruptions to both the biological material, for example mutations or changes to the hyperfine environment, as well as changes to the external field will change the balance of reactive oxygen, likely resulting in oxidative stress. As a caveat we acknowledge that this would very much depend on the exact parameters of the hyperfine coupling tensors for biologically appropriate radical pairs. What we have demonstrated in this paper is that specific choices (in our model system) for the strength and anisotropy of the hyperfine parameters can optimise specific products of spin chemistry reactions in specific magnetic fields.

\subsection{Significance for space exploration (and life on Earth)}
If terrestrial biology has evolved to be coupled to the specific strength of Earth's magnetic field, then the effects of weak magnetic fields on integral biological signalling mechanisms such as ROS have significance for a number of aspects of space exploration, particularly for the case of longer term settlement on other planets. In this instance, the new planet's magnetic field will have an effect on physiological health but also on the health of the food that must necessarily be grown for consumption. While there is some research into the role that microgravity may play in the successful propagation of plants, there is less attention paid to the role that differing or non-existent magnetic fields may play. Plant growth and robustness has been shown to respond to magnetic fields \cite{maffei,radha}, a fact that will have to be considered by future populations on new planets. Magnetic fields have also been implicated in germination processes in plants \cite{bezerra}. That magnetic fields are implicated in processes related to reproduction would also have profound consequences for the long term viability of human settlements in space. Magnetic fields play some role in embryogenesis, the full import of which remains to be revealed \cite{savic,formicki}. The complexities of magnetic field interactions in biological organisms, will likely involve more complex solutions than the simple introduction of artificial magnetic fields to space craft, for instance.\\
\\
Further research may also reveal ways in which biological systems have adapted to fluctuations in the magnetic field, evolving endogenous magnetic fields as protective mechanisms. An example of this is investigated in a paper that describes how the fast-relaxing spin of iron in photosynthetic networks can generate a magnetic field \cite{marais}. This field in turn modulates singlet-triplet conversion in radical pairs and minimises the conversion of oxygen radicals into potentially harmful products \cite{marais}. Iron is ubiquitous in biological systems. Iron is also essential to germination processes, with iron deficiency promoting seed dormancy \cite{murgia}. Iron-containing proteins such as ferritin also play a role in the management of ROS \cite{ravet}. Ferritin levels are also of importance in human embryogenesis, with low birth weight correlating with low levels of ferritin \cite{scholl}. In the context of embryogenesis and magnetic fields, the arrangement of iron-rich ferritin in the placenta is of particular interest, with one paper demonstrating that the magnetic susceptibility of ferritin iron increases dramatically when arranged within the placenta, as opposed to crude or purified ferritin extracts \cite{mykhaylyk}. Given the effects of magnetic fields on biological systems, in particular the development of stem cells, might the arrangement of iron in the placenta have evolved to generate its own stable magnetic field to mitigate any magnetic field disruption that might harm the developing embryo? There is growing interest in the importance of endogenous electric fields in generative and regenerative processes  \cite{levin2012}. It would potentially be interesting to investigate the capacity for biological systems to produce their own endogenous magnetic fields, especially given the role they play in modulating ROS.\\
\\
A better understanding of weak magnetic fields in the biological context would not only benefit the health of future settlements on other planets, but also terrestrial health. The Earth's magnetic field is itself not completely uniform. It fluctuates over time, on short and long timescales. Geomagnetic storms, for instance, have been suggested to correlate with both physiological and psychological disruptions, though far more rigorous investigation is needed \cite{close,kay,ora}. The reversal of the magnetic poles, which leads to an extended period of decreased magnetic field strength, has been suggested to coincide with mass extinctions, though this remains a hypothesis \cite{erdmann}. In addition to this, the geomagnetic field varies in magnitude at different latitudes, ranging from roughly 20 to 70 $\mu$T. If magnetic fields do play a role in the maintenance of health, it would follow that different latitudes might experience different health challenges. Given the potential role of ROS in inflammation, it might be expected that inflammatory conditions demonstrate latitude dependence. This is indeed the case, with certain bio-markers of conditions such as idiopathic inflammatory myopathies demonstrating latitude dependence, although it is not clear whether this is due to differences in UV radiation \cite{aguilar}. Inflammatory bowel disease also shows increased hospitalisation at higher latitudes in a manner that is not fully explained by differences in season or exposure to UV light \cite{stein}. The latitudinal gradient effect is also described for multiple sclerosis, with elements of the disease being strongly positively correlated with increasing latitude towards the poles, on both sides of equator \cite{sabel,simpson}. An analysis of infantile spasms also discovered a geographical difference in incidence which is consistent with a latitudinal contribution to epileptogenesis, although the study was limited by a lack of representation from the Southern hemisphere \cite{jia}. Even COVID-19 displays a latitude bias, although, as with many of the diseases listed here, the hypothesis is that this is related to levels of vitamin D \cite{whittemore}. While the cause of this latitude bias in instances of disease remains unclear, we suggest that magnetic field variations might play some role, and as such deserve closer scrutiny in the context of physiology.
\section{Conclusions}
The results of this paper demonstrate a clear dependence of radical pairs with biological parameters on an external magnetic field. Although we employ a simple toy model, we believe that it is sufficient to display the effect. More detailed approaches are necessary to fill in the exact parameters of the effect. These approaches should include theoretical modelling but perhaps more importantly, the development of sufficiently sensitive technology or sensors that might illuminate the exact interactions of magnetic fields in living systems. The model we have employed here approximates an oxygen radical. It is difficult to overstate the importance of ROS in biological systems, both in terms of signalling as well as inflammation and cell damage. A better understanding of the role that magnetic fields play in possibly modulating ROS would be beneficial for at least two reasons that we illustrate in this paper. First, the evolution of biological materials is coupled the the Earth's magnetic field, any change of this field will alter biological function. The travel and settlement of humankind outside of the Goldilocks Zone of the geomagnetic field will thus depend on an ability to predict and ameliorate the health concerns that may arise from altered magnetic fields. In the short term, however, this understanding equally applies to life on Earth. The Earth's magnetic field ranges across latitudes and, as demonstrated in this paper, different latitudes have different radical pair yields and thus a difference in ROS. This is interesting in the context of a recorded latitude dependence of a number of inflammatory diseases, from multiple sclerosis to COVID-19. While a number of other latitude dependent factors might play a role, magnetic field differences could also offer an explanation. While  microgravity conditions might be more difficult to simulate on Earth, studying the effects of magnetic fields on biological systems is implementable on Earth, through the creation and modulation of different magnetic field technologies to further our understanding of human health and disease, both now and in our multi-planetary future.

\bibliography{QEbib}



\section*{Funding and Acknowledgments}
B.A, A.H., I.S. and F.P. were supported by the National Institute for Theoretical and Computational Sciences. Thank you to Tamsyn Adams for the diagram in Figure 1. Thank you to Clarice Aiello for her comments and ideas. We also acknowledge The Guy Foundation for their support.
\end{document}